\documentclass[epj,nopacs]{svjour}
\usepackage{graphicx}

\begin{document}

\title{Improved Extrapolation Methods of Data-driven 
Background Estimations in High Energy Physics}


\author{Suyong Choi and Hayoung Oh}


\institute{Department of Physics, Korea University, Seoul 02841, Republic of Korea \\
\email{suyong@korea.ac.kr, hayoung.oh@cern.ch}
}


\abstract{
Data-driven methods of background estimations are often used to 
obtain more reliable descriptions of backgrounds. 
In hadron collider experiments, data-driven techniques are 
used to estimate backgrounds due to multi-jet events, which are difficult to model accurately.
In this article, we propose an improvement on one of the most widely 
used data-driven methods in the hadron collision environment, the ``ABCD'' method of extrapolation.
We describe the mathematical background behind the data-driven methods and extend the idea
to propose improved general methods.
}

\authorrunning{S. Choi and H. Oh}
\titlerunning{Improved Data-driven Background Estimation}

\maketitle

\section{Introduction}
The Standard Model (SM) of particle physics is compatible with almost  all of the 
measurements from particle experiments. 
In contrast to the successes on Earth, astrophysical measurements
seem to imply existence of energy component that cannot be explained by
SM, and pose a serious challenge.

Despite theoretical and experimental efforts, there is no 
direct evidence that any of the solutions proposed is correct. Moreover,
it is not clear what direction should be taken in order to resolve the problem.
Particles predicted by viable extensions of the SM are 
already excluded beyond many TeV's at the LHC \cite{exotics}.
It may turn out that these new states are massive enough to be 
beyond the reach of the LHC for direct production.
However, it does not exclude the possibility that interesting physics are
waiting to be found in rarer and more 
complicated final states. For example, we may have to
entertain the possibility of exotic final states \cite{continuum1,continuum2},
where new states appear as a continuum rather than as a resonance, above 
backgrounds. In either case, better accuracy of background 
estimation is necessary.

For many processes of interest, automatic
calculations to next-to-leading order (NLO) in strong interactions 
are accessible in modern Monte Carlo event generators \cite{mg5}. 
However, even at the NLO, theoretical uncertainties are larger
than statistical uncertainty for many processes at the LHC. 
And as the number of final-state hadronic jets increase, even the 
accuracy of NLO calculations decreases \cite{ttbbNLO}. 
Parton showering, hadronization, and underlying events
have smaller effect on the theoretical uncertainty, but nevertheless
are not negligible.

To reduce the uncertainties related to background estimation, various
data-driven estimation methods could be employed. 
Data-driven methods make use of the data in the ``background'' dominated control region (CR) to estimate 
background contributions in the ``signal'' region (SR), where interesting events may 
be found.
The method of interpolating using side-bands is a canonical method. 
In analyses involving hadron collision data, we often employ a method of extrapolation, called ``ABCD,'' a data-driven background estimation method. 
It should be noted that data-driven methods do not entirely exclude the use of simulated data.
In this article, we review the main idea behind data-driven methods and then extend
it to find an improvement for the extrapolation method. 

\section{Data-driven methods of background estimation}
The concept of estimating backgrounds from the data itself is nothing new. Important
discoveries in the history of particle physics would not have been possible without such
estimations, given that the underlying theory of particle interactions 
were not very well known or had large uncertainties \cite{jpsi1}-\cite{top2}.

While there are many ways that data-driven methods can be divided, in
this article, we will group them into two categories.
In the first category, there are data-driven methods that use interpolations
from the measurements performed on the side-bands. These methods are used
when we look for a new particle state in a restricted range of kinematic phase space (usually mass).
In the second category, there are methods we use when straight interpolations are difficult
to employ. 
The methods that use extrapolations based on information in signal-depleted regions, fall in this group. 
An extrapolation method, called the ``ABCD'' method, is often used in hadron collider experiments, 
where predictions of multijet production processes have large uncertainties 
\cite{dzero1,dzero2,matrixmethod}.
For more complicated analyses, it could involve combinations of the two categories.

\subsection{Interpolation methods}
We briefly review the interpolation methods, which will give us
ideas on how to extend and improve extrapolation methods. In an interpolation method, measurements 
are performed in the side-bands or CRs that surround the SR
and the information is combined to estimate the backgrounds 
in the signal region. In the absence of other information,
the minimal assumption is that the background would have a smooth distribution. 

Let us take a one-dimensional example. We may assume that the signal region
is in $x_0 \sim x_0+\Delta$, Without loss of generality. The number of backgrounds
in this region for a distribution of backgrounds described by $f(x)$ 
may be expressed as $F(x_0) \equiv \int_{x_0}^{x_0+\Delta} f(x) dx$. Let us take a simple side-band of
equal width to either side of the signal region. The backgrounds on the left(right) side-band are $F(x_0-\Delta)$ ($F(x_0+\Delta)$), respectively. If we assume that the series expansion 
is valid, we can then express the entries in the side-bands as
\begin{eqnarray}
F(x_0-\Delta) & = & F(x_0)-\Delta F'(x_0)+ \Delta^2 \frac{1}{2} F''(x_0) \nonumber \\
            & &    - \Delta^3 \frac{1}{3!} F'''(x_0)  + O(\Delta^4) \\
F(x_0+\Delta) & = & F(x_0)+\Delta F'(x_0)+ \Delta^2 \frac{1}{2} F''(x_0) \nonumber \\
            & &   + \Delta^3 \frac{1}{3!} F'''(x_0) + O(\Delta^4). 
\end{eqnarray}
From the two side-bands, the best estimate of $F(x_0)$ is obtained by taking the average of the two:
\begin{equation}
    F(x_0) = \frac{1}{2} \left[F(x_0-\Delta) + F(x_0+\Delta)\right]+ O(\Delta^2),
\end{equation}
which is a well-known result.

For a background whose distribution is of the $f(x)=ax+b$ form, the answer is exact. However, for a shape that has higher-order terms, this approximation may not be enough. If we allow two side-bands on each side, the terms proportional to $\Delta^2$ can be eliminated.
\begin{eqnarray}
F(x_0-2\Delta) &  = & F(x_0)-2\Delta F'(x_0)+ 2 \Delta^2 F''(x_0) \nonumber \\
 & & - \Delta^3 \frac{8}{3!} F'''(x_0)  + O(\Delta^4) \\
F(x_0+2\Delta) & = & F(x_0)+2\Delta F'(x_0)+ 2 \Delta^2 F''(x_0) \nonumber \\
 & &+ \Delta^3 \frac{8}{3!} F'''(x_0) + O(\Delta^4). 
\end{eqnarray}
The best estimate from two equal width side-bands on each side is
\begin{eqnarray}
    F(x_0) & = & \frac{4}{6} \left[F(x_0-\Delta) + F(x_0+\Delta)\right] \nonumber \\
    & & - \frac{1}{6} \left[F(x_0-2\Delta) + F(x_0+2\Delta)\right] + O(\Delta^4),
\end{eqnarray}
which is accurate for background distribution $f(x)$ that is locally a cubic function. 
One can easily understand this, since with one side-band on each side, we can fit a 
line through the two measurement points for interpolation, and
thus find the linear function exactly. 
And with two side-bands on each side, we have four measurements, therefore,
we can fit a cubic function for interpolation.

A similar idea can be adapted to a case with more than one dimension.
Let us consider a rectangular signal region in $x, y$ space between $x_0 \sim x_0+\Delta_x$ and $y_0 \sim y_0+\Delta_y$. Altogether, we can use 8 side-bands, four on the sides of the rectangle and four regions on the corners. Without any prior knowledge of the background distributions, and using similar arguments as before, the best estimate for interpolation is
\begin{eqnarray}
    F(x_0, y_0) = \frac{1}{4}  & [ & 2F(x_0-\Delta_x, y_0) + 2F(x_0+\Delta_x, y_0) \nonumber \\
    & & + 2F(x_0, y_0-\Delta_y) + 2F(x_0, y_0+\Delta_y) \nonumber \\
    & &  - F(x_0-\Delta_x, y_0-\Delta_y)  \nonumber \\
    & & - F(x_0 + \Delta_x, y_0-\Delta_y) \nonumber \\
    & &  - F(x_0-\Delta_x, y_0+\Delta_y)  \nonumber \\
    & &- F(x_0 + \Delta_x, y_0+\Delta_y) ] + O(\Delta^4).
\end{eqnarray}

\subsection{``ABCD'' extrapolation methods}
In background estimation using interpolation methods, the signal is completely surrounded by 
CRs that provide strong constraints. They would be useful if the signal is localized.
However, in searches for new physics signatures at large energies,
the signal of interest is expected to populate higher energy, mass, or jet multiplicity
regions. In these cases, measurements based on the signal-depleted CRs must be
extrapolated to the SR.

We introduce the notation to be used for the extrapolation methods.
We can use the extrapolation methods of background estimation 
if the dependence of an observable on $x$ and $y$ is mostly independent, as:
\begin{equation}
    P(x,y) = P_x(x) P_y(y)\left[1+\epsilon(x,y)\right],
\label{eq:factorizable}
\end{equation}
where the non-independent component is in $\epsilon$.
We assume that the non-independent part is small $|\epsilon|<<1$.
Then the integral in a rectangular region would be mostly factorizable as well.
\begin{eqnarray}
    & & F(x_0,x_1,y_0,y_1) \nonumber \\
    & = & \int_{y_0}^{y_1}\int_{x_0}^{x_1} P_x(x)P_y(y) \left[1+\epsilon(x,y)\right] dx dy \nonumber \\
    & = & \int_{x_0}^{x_1} P_x(x) dx \int_{y_0}^{y_1} P_y(y) dy \nonumber \\
    & & \times \left[ 1 + \frac{\int_{y_0}^{y_1}\int_{x_0}^{x_1} P_x(x)P_y(y) \epsilon(x,y) dx dy}{ \int_{x_0}^{x_1} P_x(x) dx \int_{y_0}^{y_1} P_y(y) dy   } \right] \nonumber \\
    & = & S_x(x_0, x_1) S_y(y_0,y_1) \left[ 1 + \Sigma(x_0, x_1, y_0, y_1) \right],
\end{eqnarray}
where $\Sigma$ is the average value of $\epsilon$ over this range and depends on
the amount of dependence between the two variables, $x$ and $y$.
$S_x$($S_y$) is the integral of $P_x$($P_y$) in the range $x_0\sim x_1$ ($y_0\sim y_1$), respectively.
For a fixed-width window, $x_1=x_0+\Delta_x$ and $y_1=y_0+\Delta_y$, $F$ is a function of 
$x_0$ and $y_0$, so we can omit the arguments $x_1$ and $y_1$ as
\begin{equation}
    F(x,y) = S_x(x) S_y(y) \left[ 1+\Sigma(x,y)\right].
    \label{eq:F}
\end{equation}

An estimate of $F(x,y)$ is obtained by taking suitable products of the $
F$s in the neighboring regions as:
\begin{eqnarray}
\label{eq:extrap}
    & & \frac{F(x-\Delta_x, y)F(x, y-\Delta_y)}{F(x-\Delta_x, y-\Delta_y)} \nonumber \\
    & = & S_x(x)S_y(y)\left[1+\Sigma+ \left(\frac{1}{1+\Sigma}\frac{\partial\Sigma}{\partial x}\frac{\partial\Sigma}{\partial y} - \frac{\partial^2\Sigma}{\partial x \partial y}\right)\Delta_x\Delta_y \right] \nonumber \\ 
    & & + O(\Delta^3) \nonumber \\ 
    & = & F(x,y) + O(\Delta^2),
\end{eqnarray}
where the $\Delta$'s stand for either $\Delta_x$ or $\Delta_y$. 
The $\Delta_x\Delta_y$ term would vanish if $\epsilon(x,y)\rightarrow 0$. 
Therefore, the error of the estimation depends on the degree of non-independence of $x$ and $y$.
In this derivation, we do not assume that $S_x$ ($S_y$) vary slowly as a function of $x$ ($y$), respectively, but that $\Sigma$ varies slowly enough that the series expansion is valid.

The method is often referred to as the ``ABCD'' method  (Eq. \ref{eq:extrap}) or
matrix method. In an ABCD method, two-dimensional phase space is divided into
four regions, one of which is the SR and the neighboring three regions
are the CRs. The choice of the two control variables
used for this purpose depends on the physics case of interest, but should be
as independent as possible.
In hadron collision experiments, such extrapolation methods are used to estimate the
backgrounds in a variety of settings.
Usually, the signature of interest is expected  at high energies or
large particle multiplicities, therefore, the interpolation methods
cannot be used. It is in this regime where the need for these methods 
arises because of large theoretical or experimental uncertainties
in prediction using simulations or calculations.
The data-driven approach can bypass many of these difficulties.

\begin{figure}
    \centering
    \includegraphics[width=0.3\textwidth]{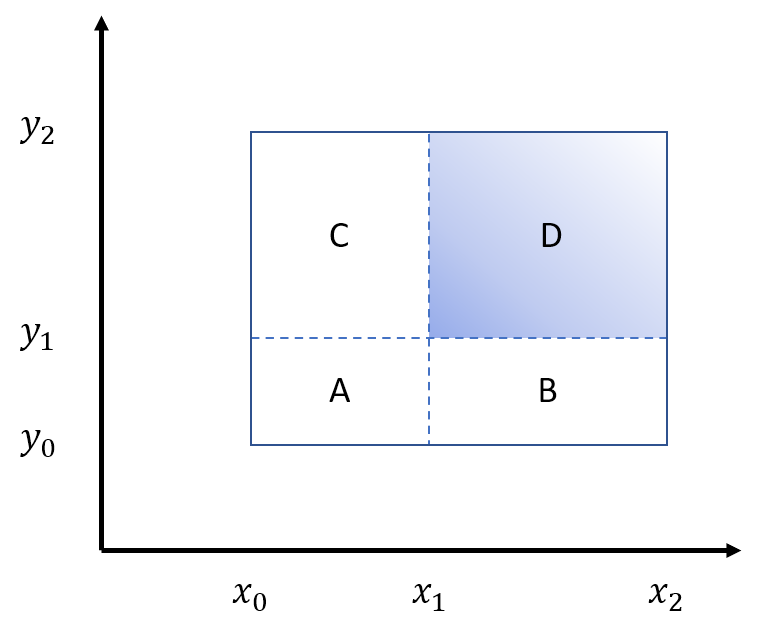}
    \caption{The various control regions and the signal region ($D$) of the ABCD method.}
    \label{fig:ABCD}
\end{figure}

The information from the three $A$, $B$, and $C$ CRs, is used to estimate the backgrounds in the signal region, $D$ (Fig. \ref{fig:ABCD}). Generally, we can express the estimate of $F_D$ as $\hat{F}_D$,
\begin{eqnarray}
\label{eq:ABCD}
    \hat{F}_D &  = & \frac{F_C}{F_A}\times F_B \nonumber \\
    & = & \frac{S_x(x_0,x_1) S_y(y_1, y_2)[1+\Sigma(x_0,x_1,y_1,y_2)]}{S_x(x_0,x_1) S_y(y_0, y_1)[1+\Sigma(x_0,x_1,y_0,y_1)]} \nonumber \\
    & & \times S_x(x_1,x_2) S_y(y_0, y_1)[1+\Sigma(x_1,x_2,y_0,y_1)] \nonumber \\
    & = & S_x(x_1,x_2)  S_y(y_1, y_2) \left[  1 + \Sigma(x_1, x_2, y_1, y_2)  \right] \nonumber \\
    & & + O(\Delta^2),
\end{eqnarray}
where the $\Delta$'s are either $x_1-x_0$, $x_2-x_1$, $y_1-y_0$, or $y_2-y_1$. 

When $x_2$ and/or $y_2$ is taken to infinity, the expansion, in general, is not valid
unless $\Sigma=0$ since $\Delta\rightarrow\infty$. However, even if $\Sigma\neq 0$, under certain conditions, the expansion could still be valid. 
For the case $x_2\rightarrow\infty$, if the distribution $P_x(x)$ falls sharply as $x$ increases, then 
 Eq. \ref{eq:ABCD} could be still valid.
Since $\Sigma(x_1, x_2, y_0, y_1)\approx \Sigma(x_1, x_1+\delta_x, y_0, y_1)$, remembering that
 $\Sigma$ is the average value of $\epsilon$ in the given region, thus $x_2$ is 
not as relevant since the data are distributed heavily towards lower values of $x$. 
Under these conditions,
\begin{eqnarray}
    & &  \frac{1+\Sigma(x_0,x_1,y_1,y_2)}{1+\Sigma(x_0,x_1,y_0,y_1)} \times [1+\Sigma(x_1,x_2,y_0,y_1)] \nonumber \\
    &= & 1  +\Sigma(x_1,x_2,y_0,y_1) + \Delta_{y1} \Sigma_3(x_0, x_1, y_1, y_2) \nonumber \\
    & & + \Delta_{y2} \Sigma_4(x_0, x_1, y_1, y_2) + O(\Delta_y^2) \nonumber \\
    &\approx & 1 + \Sigma(x_1,x_2,y_0,y_1) + \Delta_{y1} \Sigma_3(x_0, x_0+\delta, y_1, y_2) \nonumber \\
    & & + \Delta_{y2} \Sigma_4(x_0, x_0+\delta, y_1, y_2) + O(\Delta_y^2)\nonumber \\
    &\approx & 1 +  \Sigma(x_1,x_2,y_0,y_1) + \Delta_{y1} \Sigma_3(x_1, x_1+\delta, y_1, y_2) \nonumber \\
    & & + \Delta_{y2} \Sigma_4(x_1, x_1+\delta, y_1, y_2) -  \nonumber \\
    & & \Delta_{x1}\Delta_{y1} \Sigma_{31}(x_1, x_1+\delta, y_1, y_2) \nonumber \\
    & & - \Delta_{x1}\Delta_{y2} \Sigma_{41}(x_1, x_1+\delta, y_1, y_2) + O(\Delta_y^2)\nonumber \\
    & \approx & 1 + \Sigma(x_1, x_2, y_1, y_2) + O(\Delta^2),
\end{eqnarray}
where $\Sigma_i$ ($\Sigma_{ij}$) is the partial derivative with respect to the $i$th argument ($i$ and $j$ arguments), respectively, and $\Delta$s are either $\Delta_{x1}$, $\Delta_{y1}$, or $\Delta_{y2}$.
In summary, with the ABCD method, measurements in three regions neighboring the 
SR can be used to give the accurate description to $O(\Delta^2)$, given
that the correlation between the $x$ and $y$ is weak and the distribution falls sharply
in $x$ and $y$.

\section{Improving the data-driven extrapolation method}
As was the case with interpolation, it is possible to improve the accuracy of extrapolation methods by including more CRs. We derive several new analytic results and
provide some case studies to demonstrate their efficacy.

\subsection{Extended ABCD methods}

\begin{figure}
    \centering
    \includegraphics[width=0.3\textwidth]{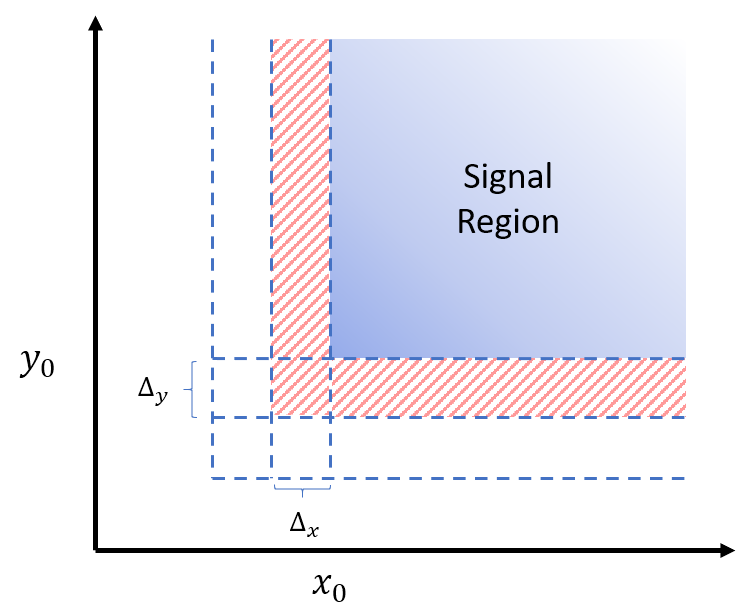}
    \caption{Control regions used in the extended ABCD methods. The upper right region is the signal region, while the rest are the control regions. The hatched regions are the nominal regions used
    in an ABCD method, while the other open regions (in addition to the hatched regions) are incorporated in the 
    extended ABCD methods.}
    \label{fig:ABCDext}
\end{figure}
We assume that the SR is $x>x_0$ and $y>y_0$ (Fig. \ref{fig:ABCDext}) and that the joint distribution in $x$ and $y$ is mostly factorizable. Then we can express the number of entries in the SR as $F(x_0, y_0) = S_x(x_0)S_y(y_0)[1+\Sigma(x_0, y_0)]$. By using more information in the 
CRs $[x_0-2\Delta_x, x_0-\Delta_x]$  as well as $[x_0-\Delta_x, x_0]$ and similarly in $y$,
the accuracy can be improved as
\begin{eqnarray}
    \label{eq:ABCDextended2bins}
    & & F(x_0, y_0) \nonumber \\
    & = & \left[ \frac{F(x_0-2\Delta_x, y)  F(x_0, y_0-2\Delta_y)}{F(x_0-2\Delta_x, y_0-2\Delta_y)} \right]^{-\frac{1}{3}} \nonumber \\
    & &  \cdot \left[    \frac{F(x_0-\Delta_x, y_0)  F(x_0, y_0-\Delta_y)}{F(x_0-\Delta_x, y_0-\Delta_y)}  \right]^{\frac{4}{3}} + O(\Delta^3),
\end{eqnarray}
where $\Delta$ stands for either $\Delta_x$ or $\Delta_y$. 
With fixed-width CRs, terms up to $\Delta^2$ can be exactly canceled. Therefore, the
effects of correlations among variables on the prediction are mitigated as well.
In the appendix, we give an explicit expression for Eq. \ref{eq:ABCDextended2bins}.

We can extend the idea further by using information in 
eight CRs (Fig. \ref{fig:ABCDext}), where it is possible to get accuracy of the $O(\Delta^4)$ order:
\begin{eqnarray}
    & & F(x_0, y_0) \nonumber \\
    & = & \frac{F(x_0-2\Delta_x, y_0) F(x_0, y_0-2\Delta_y)}{F(x_0-2\Delta_x, y_0-2\Delta_y)} \nonumber \\
    & & \cdot \left[ \frac{F(x_0-\Delta_x, y_0) F(x_0, y_0-\Delta_y)}{F(x_0-\Delta_x, y_0-\Delta_y)} \right]^4 \nonumber \\
    & & \cdot \left[ \frac{F(x_0-2\Delta_x, y_0) F(x_0, y_0-\Delta_y)}{F(x_0-2\Delta_x, y_0-\Delta_y)} \right]^{-2} \nonumber \\
    & & \cdot \left[ \frac{F(x_0-\Delta_x, y_0) F(x_0, y_0-2\Delta_y)}{F(x_0-\Delta_x, y_0-2\Delta_y)} \right]^{-2} + O(\Delta^4).
    \label{eq:ABCDextended_8regions}
\end{eqnarray}

However, having more CRs does not always result in reduced error.
Since the method involves multiplication or division operations, statistical uncertainties,
due to the finite number of entries in each CR directly affect the uncertainty of the prediction. 
From practical considerations, it may be desirable to have fewer CRs,
so we also derived an optimal expression for the case of five control regions, by
allowing for two control region bins in either $x$ or $y$, but not in both.
In the case of two control region bins in $x$, but one in $y$,  the optimal combination
of the control region measurements is
\begin{eqnarray}
    & & F(x_0, y_0) \nonumber \\
    & = & \left[\frac{F(x_0-\Delta_x, y_0)  F(x_0, y_0-\Delta_y)}{F(x_0-\Delta_x, y_0-\Delta_y)}
    \right]^2 \nonumber \\
    & & \cdot 
    \left[ \frac{F(x_0-2\Delta_x, y_0-\Delta_y) }{F(x_0-2\Delta_x, y_0) F(x_0, y_0-\Delta_y) } \right] \nonumber \\
    & & + O(\Delta_x^2\Delta_y).
    \label{eq:extendedABCD_5regions}
\end{eqnarray}
As before, the error depends on the assumptions of
weak correlations among the dependent variables $x$ and $y$, as described
by $\epsilon(x,y)$. We also assume that $\epsilon(x,y)$ varies slowly enough
to allow for the series expansion.

While the results derived are for fixed width bins, they can be applied to the 
variable widths cases. The variable-widths bins could be modified into fixed-width bins by locally
stretching or squeezing the control variables phase space. And as long as this operation
does not invalidate the assumption of the weak correlations, these methods are 
applicable.

\subsection{Case studies of extended ABCD methods}
\subsubsection{Toy example}

\begin{figure}
    \centering
    \includegraphics[width=0.45\textwidth]{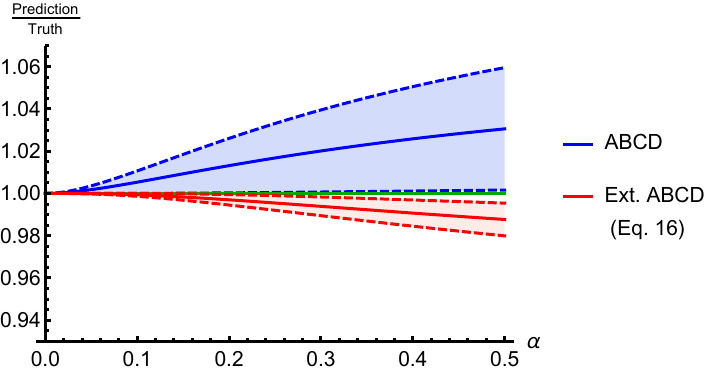}
    \caption{Plot of ratio of prediction to the truth of the different extrapolation methods as a function
    of $\alpha$ together with error bands for the example distribution in Eq. \ref{eq:toy}.}
    \label{fig:closure}
\end{figure}

As a simple test, we apply the ABCD method and the extended ABCD method of Eq. \ref{eq:extendedABCD_5regions} 
to a distribution
\begin{equation}
    \frac{1}{1+\frac{1}{2} x^2} \frac{1}{1+ y^2}\left[1+\alpha(x+y)\right],
    \label{eq:toy}
\end{equation}
which is a smoothly decreasing distribution in $x$ and $y$, but otherwise arbitrary.
The distribution would separable in $x$ and $y$ in the absence of the $x+y$ term, which
provides some correlation between $x$ and $y$.
For simplicity, the boundaries for the ABCD method are set to $x_0=1$, $x_1=2$, $x_2=3$, $y_0=1$, $y_1=1$, $y_2=2$.
The true value of the area in $D$ is $F_D=0.1210$ for $\alpha=0.5$, while the ABCD method (Eq. \ref{eq:ABCD}) yields $0.1247$.
The extended ABCD method with the left boundary at $x_{-1}=0$ yields $0.1195$. Extended ABCD method reduces the error in prediction by a factor of 2.5 for this case. 

Fig. \ref{fig:closure} shows how the predictions of ABCD and extended ABCD change with $\alpha$.
The bands represent the error terms of the respective methods in the appendix. Since the distribution
is known explicitly,  the error terms can be calculated.
As $\alpha\rightarrow0$, both methods converge to 1, as expected, since the distribution becomes independent in $x$ and $y$.

\subsubsection{$t\bar{t}$+multi-jets in hadronic channels}
For the second case study, we apply various ABCD methods of background estimations to $t\bar{t}+jj$ simulated sample.
The $t\bar{t}$+multi-jets processes are backgrounds to many of the searches for 
physics beyond the standard model at the LHC \cite{atlassusy,cmsfourtop}.
While calculation of $t\bar{t}+jj$ is available at the next-to-leading order (NLO), 
it has relatively larger theoretical uncertainties than what is desired by the experiments \cite{ttbbNLO}. 
Furthermore, the quoted uncertainties in the literature are on the overall inclusive cross sections, but in some phase space, 
the uncertainties on the the differential cross sections could be even larger.
It is difficult to envision improved  calculations for these processes 
in the foreseeable future.
Therefore, having a more reliable data-driven technique is important for these processes.

We generated one million events of $pp\rightarrow t\bar{t}jj$ sample at $\sqrt{s}=14$~TeV with MG5aMC@NLO v2.61 at LO \cite{mg5}. The extra partons are required to have $p_T>20$~GeV and $|\eta|<5.0$. The partons are hadronized with Pythia 8 \cite{pythia8}.
Delphes 3 fast detector simulation and reconstruction were subsequently applied.
The reconstructed jets are required to be $p_T>30$~GeV and $|\eta|<2.4$.
We required zero isolated lepton that satisfies $p_T>20$~GeV and $|\eta|<2.4$ in an event.

\begin{figure}[t]
    \centering
    \includegraphics[width=0.30\textwidth]{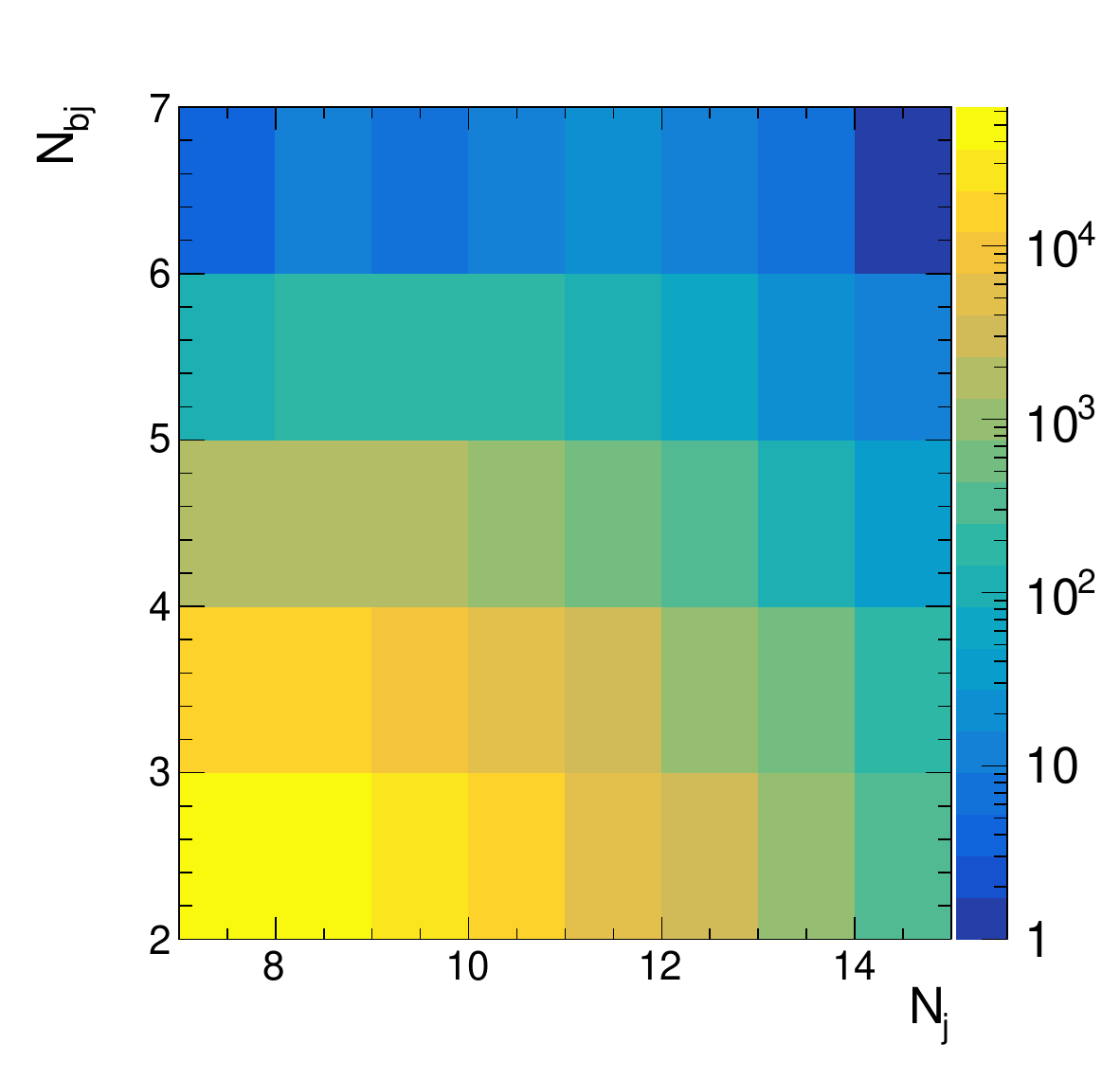}
    \caption{Distribution of the number of jets ($N_j$) and the number of b-tagged jets ($N_{bj}$) in $t\bar{t}+multijets$ sample. 
    }
    \label{fig:nbnj}
\end{figure}

\begin{table}
    \centering
\begin{tabular}{c|ccc}
$N_{bj}$ & \multicolumn{3}{c@{}}{$N_j$}\\
    & 7  & 8 & $\geq 9$  \\
   \hline
  2   & 63216  & 49685 & 55756 \\
  3 & 15046 & 14378 & 20068 \\
  $\geq4$ & 1961 & 2388 & 4874

\end{tabular}
    \caption{Number of events in various $N_{bj}$ and $N_j$ regions in $t\bar{t}+multijets$ samples. The SR considered in this study requires $N_j\geq 9$ and $N_b\geq4$.}
    \label{tab:njnb}
\end{table}

\begin{table}
    \centering
    \begin{tabular}{c|c c}
      Extrapolation method   & Prediction ($\hat{F}_D$) & $\hat{F}_D/F_D$ \\
      \hline
      ABCD (Eq. \ref{eq:ABCD}) & $3333\pm 77$ &  $0.684\pm 0.015$ \\
      Ext. ABCD (Eq. \ref{eq:ABCDextended2bins}) & $4149 \pm 132$ &  $0.851\pm 0.027$\\
      Ext. ABCD (Eq. \ref{eq:ABCDextended_8regions}) & $4352\pm 271$ & $0.893\pm 0.056$\\
      Ext. ABCD (Eq. \ref{eq:extendedABCD_5regions}) & $4247\pm 217$ & $0.871\pm 0.045$
    \end{tabular}
    \caption{Predictions of the number of events for $N_j\geq 9$ and $N_{bj}\geq 4$ in $t\bar{t}+multijets$ samples
    using various extrapolation methods and ratios with respect to the true value (4874). The statistical uncertainties
    on the predictions are calculated from Poisson fluctuations of the control regions.}
    \label{tab:abcdpredictions}
\end{table}

The distribution of the number of hadronic jets ($N_j$) and the number of $b$-tagged
jets ($N_{bj}$) is shown in Fig. \ref{fig:nbnj}, and the number of entries in each bin is listed
in Table~\ref{tab:njnb}. The correlation coefficient of the two variables is 0.139, hence,
they are weakly correlated.
We apply the methods in Eqs. \ref{eq:ABCDextended2bins}-\ref{eq:extendedABCD_5regions}, taking $N_j$ and $N_{bj}$ as control variables. 
The SR is $N_j\geq 9$ and $N_{bj}\geq 4$.
It could be applicable in a scenario where signature of interest consists of multijets and multiple $b$-tagged jets.

The results of applying various extrapolation methods are shown in Table \ref{tab:abcdpredictions}.
The uncertainties in the predictions are statistical uncertainties due to the number of entries in the
control region. They are evaluated by an ensemble test where the number of entries in
each control region fluctuates according to a Poisson distribution.
The extended ABCD methods allow for better prediction in terms of reduced deviation from the truth,
at the cost of increased statistical uncertainties.

\begin{figure}[t]
    \centering
    \includegraphics[width=0.24\textwidth]{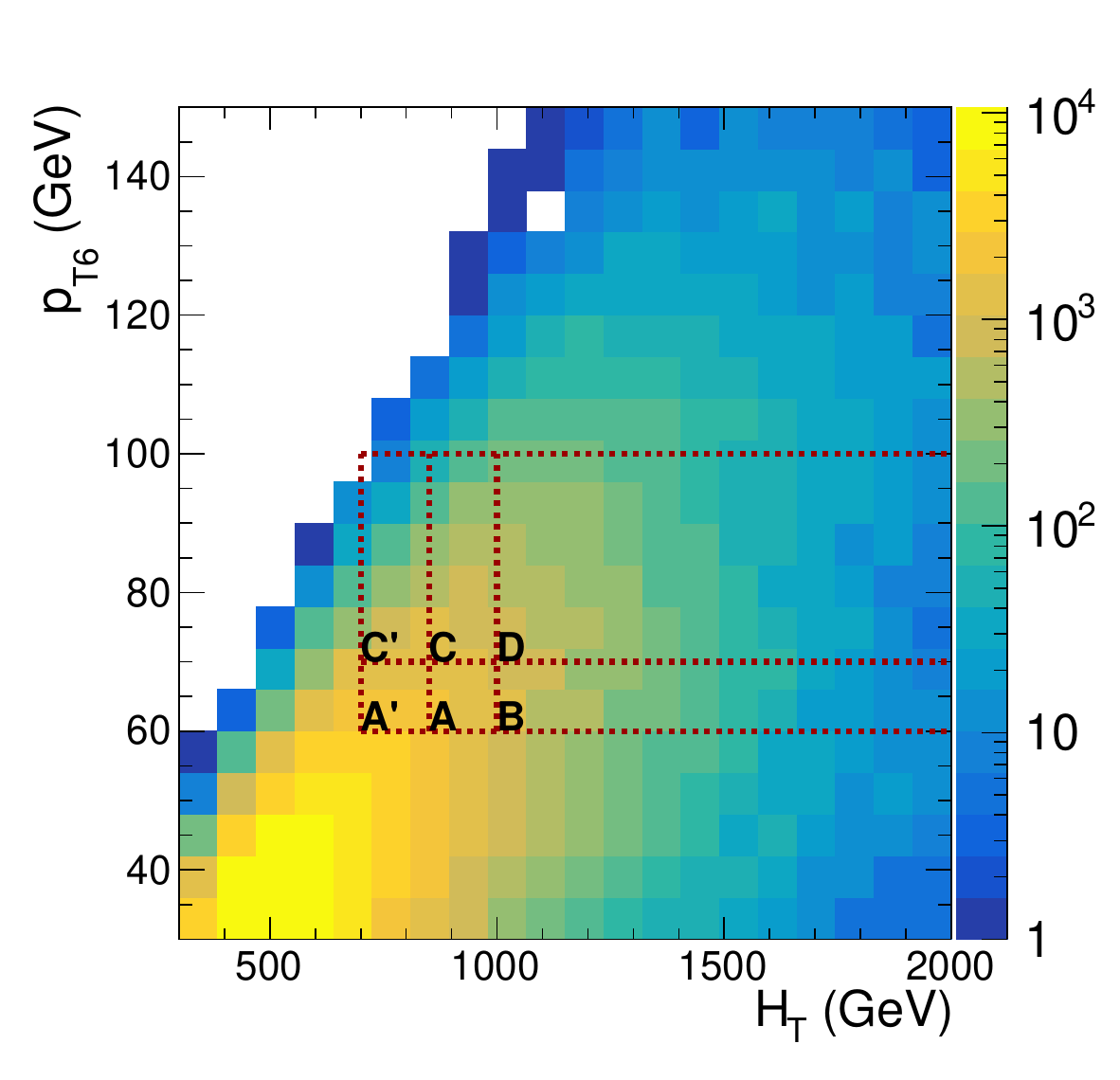} \includegraphics[width=0.24\textwidth]{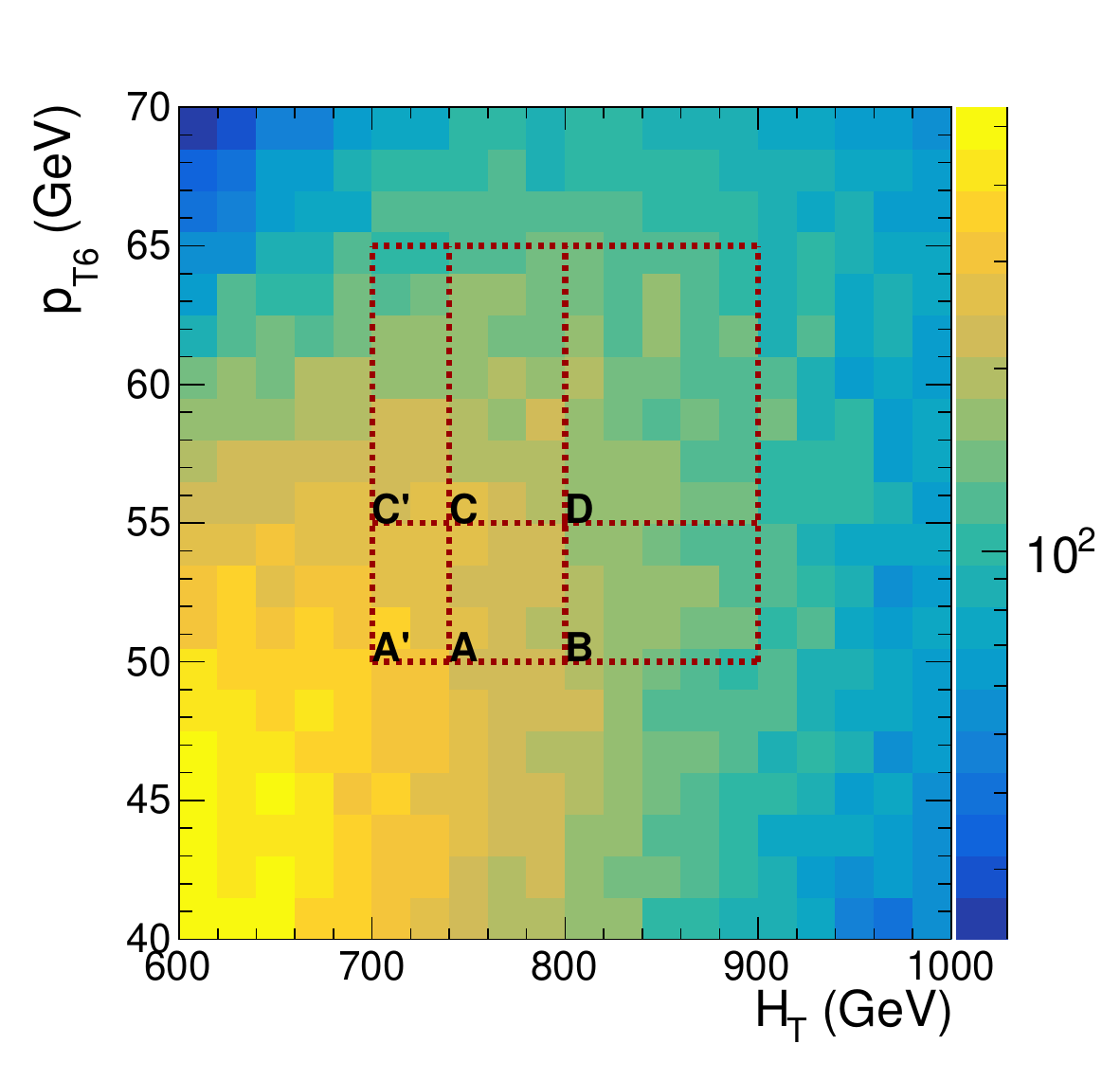}
    \caption{Distributions of $p_{T6}$ versus $H_T$ in $t\bar{t}jj$ events for the two cases of 
    applications of extended ABCD methods. Note the different scales on the axes. 
    Various control and signal regions are delineated. }
    \label{fig:htpt6}
\end{figure}

Next, we consider cases where the control variables are continuous. 
We take the hadronic scalar sum of jet transverse momenta ($H_T$) 
and the sixth leading jet transverse momentum ($p_{T6}$) as the control variables. The two variables are
obviously correlated (correlation coefficient: 0.660), as shown in Fig. \ref{fig:htpt6}. We deliberately
chose these variables to better exemplify the advantages of the extended ABCD
methods.

Since the distribution drops rapidly as $H_T$ or $p_{T6}$, we consider two different
use cases. In the first case, the widths of the CRs and SR ($\Delta_x$) are wider than the widths of the distribution,
and in the second case, the widths are similar or smaller than the width of the distribution of each control variable (Fig. \ref{fig:htpt6}).
Table \ref{tab:htpt6} shows how the different regions are defined and the number of entries in
the respective regions for the two cases. In the first case, the region of interest ($D$) has a
lower limit on $H_T$. This could be a typical use case in hadron colliders where we are
interested in phenomena at high energies. In the second case, $D$ is much narrower, and
although this is not the most general use case, it is nonetheless interesting for illustration purposes.
The bins are chosen such that the number of entries do not vary greatly among the different regions.

In the first case, the ABCD method yields $4802\pm 122$ while the extended ABCD method of Eq. \ref{eq:extendedABCD_5regions}
yields $9976\pm 488$. The ABCD method is inadequate because of the correlation
between $p_{T6}$ and $H_T$.  In the second case, the ABCD method yields
$3886\pm 128$ while the extended ABCD method yields $4493\pm 291$.
In both cases, the presence of $A'$ and $C'$ control regions provides an additional 
lever arm and allows us to take into consideration the dependence on $H_T$ better.

\begin{table}
    \centering
\begin{tabular}{c|ccc}
\multicolumn{4}{c@{}}{Case 1}\\
\hline
$p_{T6}$ & \multicolumn{3}{c@{}}{$H_T$ (GeV)}\\
  (GeV)  & $700 - 850$  & $850 -1000$ & $>1000$  \\
   \hline
  $60-70$   & 6319 ($A'$) & 4479 ($A$) & 4343 ($B$) \\
  $70-100$   & 3364 ($C'$) & 4953 ($C$) & 9288 ($D$) \\
\end{tabular}

\begin{tabular}{c|ccc}
\multicolumn{4}{c@{}}{Case 2}\\
\hline
$p_{T6}$ & \multicolumn{3}{c@{}}{$H_T$ (GeV)}\\
  (GeV)  & $700 - 740$  & $740 -800$ & $800-900$  \\
   \hline
  $50-55$   & 1901 ($A'$) & 2332 ($A$) & 2574 ($B$) \\
  $55-60$   & 2482 ($C'$) & 3521 ($C$) & 4688 ($D$) \\
\end{tabular}

    \caption{Number of entries in various $H_T$ and $p_{T6}$ regions in $t\bar{t}+multi-jets$ samples
    for the two cases considered in Fig. \ref{fig:htpt6}. The label beside each entry indicates the region each entry corresponds to.}
    \label{tab:htpt6}
\end{table}

\begin{table}
    \centering
    \begin{tabular}{c|cc|c}
         & ABCD & Ext. ABCD & Truth \\
         \hline
         Case 1 & $4802\pm 122$ & $9976\pm 488$  & 9288 \\
         Case 2 & $3886\pm 128$ & $4493\pm 291$ & 4688
    \end{tabular}
    \caption{Predictions of entries in region $D$ for the two cases in Table \ref{tab:htpt6}. The errors quoted are the expected statistical uncertainties from pseudo-experiments. }
    \label{tab:results}
\end{table}

One of the important reasons to use the data-driven method is to reduce some of the systematic uncertainties.
Through several case studies, we demonstrate that the extended ABCD methods provide estimates that are 
closer to the truth. For cases where independent variables are not easy to find, the extended ABCD method could still take into account some of the correlations. 
In many analyses, the normalization of the background is treated as a nuisance parameter
to be constrained further by fitting to data. The extended ABCD methods can provide smaller uncertainty
on the prior of the normalization and thus move towards reducing systematic uncertainties.

\section{Conclusions}
We propose extensions to the ABCD method of extrapolated background estimation by exploiting
information from additional control regions. 
The extended ABCD methods could be useful when the control variables are not exactly independent, since 
they can mitigate the effects of correlations among the variables. Through several case studies,
we demonstrate that they provide more accurate predictions at the cost of increased statistical
uncertainties.

\begin{acknowledgement}
This work was supported in part by the Korean National Research Foundation (NRF) grants NRF-2018R1A2B6005043
and NRF-2020R1A2B5B02001726.
\end{acknowledgement}

\appendix
\section{Expressions for the extended ABCD methods}
We give an explicit expression for Eq. \ref{eq:ABCDextended2bins} up to $\Delta^3$:

\begin{eqnarray}
& & S_x(x) S_y(y)  \times  \Bigl\{1+\Sigma  \nonumber \\
& &  +\frac{2 \Delta _x \Delta _y^2}{3 (1+\Sigma )^2}  \left[-2 (\Sigma ^{(0,1)})^2 \Sigma^{(1,0)} \right. \nonumber \\
& &   \left. +2 (1+\Sigma ) \Sigma ^{(0,1)} \Sigma ^{(1,1)} \right.  \nonumber \\
& &   \left. +(1+\Sigma ) \left(\Sigma ^{(0,2)} \Sigma
   ^{(1,0)}-(1+\Sigma ) \Sigma ^{(1,2)}\right)\right] \nonumber \\ 
& & +\frac{2 \Delta _y \Delta _x^2}{3 (1+\Sigma )^2}  \left[-2 (\Sigma ^{(1,0)})^2 \Sigma
   ^{(0,1)} \right. \nonumber \\
   & & \left. +2 (1+\Sigma ) \Sigma ^{(1,0)} \Sigma ^{(1,1)} \right. \nonumber \\
& &   \left. +(1+\Sigma ) \left(\Sigma ^{(2,0)} \Sigma
   ^{(0,1)}-(1+\Sigma ) \Sigma ^{(2,1)}\right)\right]\Bigr\} \nonumber \\
&  & + O(\Delta^4)
\end{eqnarray}
To reduce clutter, we omit the arguments $(x,y)$ to $\Sigma$ function. The superscripts $(m,n)$ stand for
partial derivatives, as $\Sigma^{(m,n)} = (\frac{\partial}{\partial x})^m (\frac{\partial}{\partial y})^n\Sigma(x,y)$.

And the expression for Eq. \ref{eq:extendedABCD_5regions} up to $\Delta^3$ order is
\begin{eqnarray}
& & S_x(x)S_y(y) \times\Bigl\{1 + \Sigma \nonumber \\
& & +  \frac{\Delta _x^2 \Delta _y}{(1+\Sigma )^2} \left[\Sigma ^{(0,1)} \left((\Sigma +1)
   \Sigma ^{(2,0)}-2 (\Sigma ^{(1,0)})^2\right)\right. \nonumber \\ 
  & & \left.  +(1+\Sigma) \left(2 \Sigma ^{(1,0)} \Sigma
   ^{(1,1)}-(1+ \Sigma) \Sigma ^{(2,1)}\right)\right]\Bigr\} \nonumber \\
  & & + O(\Delta^4).
\end{eqnarray}


\end{document}